\documentstyle[epsfig,indentfirst,bm,12pt]{article}
\begin{document}
\title{\Large{\bf{Nuclear Effects in Deep
Inelastic Scattering of Charged-Current Neutrino off Nuclear
Target}}}
\author{Duan ChunGui$^{1,2,3,4}$ \footnote{\tt{ E-mail:duancg$@$mail.hebtu.edu.cn}},
Li GuangLie$^{1,3}$, Shen PengNian$^{3,1,5,6}$}

\date{}

\maketitle \noindent
{\small 1.Institute of High Energy Physics, CAS, P.O.Box 918(4), Beijing 100049, China}\\
{\small 2.Department of Physics, Hebei Normal University,Shijiazhuang,050016, China}\\
{\small 3.CCAST(World Laboratory), P.O.Box8730, Beijing 100080, China}\\
{\small 4.Graduate School of the Chinese Academy of Sciences, Beijing 100049, China}\\
{\small 5.Institute of Theoretical Physics, Chinese Academy of Sciences, P.O.Box 2735,
 Beijing 100080, China}\\
{\small 6.Center of Theoretical Nuclear Physics, National Laboratory
of Heavy Ion Accelerator, Lanzhou 730000, China}

\baselineskip 9mm \vskip 0.5cm
\begin{abstract}
Nuclear effect in the neutrino-nucleus charged-Current inelastic
scattering process is studied by  analyzing the CCFR and NuTeV data.
Structure functions $F_2(x,Q^2)$ and $xF_3(x,Q^2)$ as well as
differential cross sections are calculated by using CTEQ parton
distribution functions and EKRS and HKN nuclear parton distribution
functions, and compared with the CCFR and NuTeV data. It is found
that the corrections of nuclear effect to the differential cross
section for the charged-current anti-neutrino scattering on nucleus
are negligible, the EMC effect exists in the neutrino structure
function $F_2(x,Q^2)$ in the large $x$ region, the shadowing and
anti-shadowing effect occurs in the distribution functions of
valence quarks in the small and medium $x$ region,respectively. It
is also found that shadowing effects on $F_2(x,Q^2)$ in the small
$x$ region in the neutrino-nucleus and the charged-lepton-nucleus
deep inelastic scattering processes are different. It is clear that
the neutrino-nucleus deep inelastic scattering data should further
be employed in restricting nuclear parton distributions.



\end{abstract}

\vskip 0.5cm

\noindent {\bf I Introduction }

In the past three decades, the quark and gluon distributions in
hadrons and nuclei have been one of the most active frontiers in
nuclear physics and particle physics. The nuclear parton
distribution directly affects the interpretation of the data
collected from the nuclear reactions at high energies, for example
the nucleus-nucleus and the proton-nucleus interactions at
RHIC$^{[1]}$ and LHC$^{[2]}$. Considering the nuclear effect caused
modifications on the parton distribution function should be an
essential step for understanding the suppression of $J/\psi$
production which might be a signal of the quark-gluon plasma (QGP)
in the relativistic heavy ion collision. Precisely modified nuclear
parton distributions would especially be important in determining
the electro-weak parameters, neutrino masses and mixing angles in
neutrino physics.

\vspace{0.5cm}

In 1982, the European Muon Collaboration (EMC) reported that the
measured ratio of nuclear structure functions for the heavy (iron)
and light (deuteron) nuclei in the processes of the deep inelastic
scattering (DIS) of muon off nucleus$^{[3]}$ is significantly
different from the theoretically predicted value$^{[4]}$. That was
the first clear evidence for the nuclear effect in nuclear structure
functions, and later was called EMC effect. In fact, the EMC effect
states that, in the parton point of view, quark distributions in
bound nucleon are different from those in free nucleon. The
discovery of EMC effect triggered further studies on the sizable
nuclear effect through the DIS of muon and electron off nucleus
$^{[5-8]}$. The abundant charged-lepton DIS data showed that there
are four types of nuclear effects, shadowing effect, anti-shadowing
effect, EMC effect and Fermi motion effect, appeared in the regions
of $ x < 0.1$, $0.1 < x < 0.3$, $0.3 < x <0.7$ and $x> 0.7$, where
$x$ denotes the Bjorken variable, respectively.

\vspace{0.5cm}

Like the charged-lepton DIS, the deep inelastic neutrino scattering
is also an important process for investigating the structures of
hadrons and nuclei. In this process, the structure functions
$F_2(x,Q^2)$ and the parity-violating structure function
$xF_3(x,Q^2)$ can simultaneously be measured. In 1984, Big European
Bubble Chamber Collaboration (BEBC) published the antineutrino-
neon/deuterium DIS data in the kinematic region of $0<x<0.7$ and
$0.25<Q^2<26GeV^2$ $^{[9]}$. Their measured differential cross
section ratio in the high $Q^2$ and $0.3<x<0.6$ region$^{[9]}$ is
compatible with the muon and electron scattering data from EMC and
SLAC. In the same year, CERN-Dortmund-Heidelberg-Saclay
Collaboration (CDHS) measured events originating in a tank of liquid
hydrogen and in the iron of detector in the 400 GeV neutrino
wide-band beam of the CERN Super Proton Synchrotron(SPS)$^{[10]}$.
Comparing the measured total cross sections, differential cross
sections and structure functions for hydrogen with those for iron,
no significant difference between the structure functions for proton
and iron was observed. One year later, E545 Collaboration at
Fermilab $^{[11]}$ measured the cross sections in the deep inelastic
neutrino scattering on neon or deuterium once more. Unfortunately,
they were not able to give a definite conclusion due to substantial
statistical uncertainties. In 1987, WA25 and WA29 collaborations
studied the nucleon structure functions taken from the neutrino and
antineutrino experiments for neon and deuterium$^{[12]}$. The
combined neutrino and antineutrino differential cross section data
also showed that the cross section ratios between the heavy targets
and deuterium decrease when x increases from 0.2 to 0.6, which is
again the EMC effect. In fact, many neutrino DIS experiments were
carried out with their own primary physical goals, for instance the
structure of proton, the mixing angles of electro-weak interaction
and etc., but none of them can individually confirm the EMC effect.

\vspace{0.5cm}

In early 1960's, Adler$^{[13]}$ proved that in the $Q^2\rightarrow
0$ limit, the structure function $F_2(x,Q^2)$ obtained form the
charged lepton DIS process should go to zero, but $F_2(x,Q^2)$ form
the neutrino DIS process should approach to a positive constant.
This discrepancy is caused by the the partial conservation of axial
currents(PCAC) in the weak interaction. With the aid of Adler's
theorem, J.S.Bell$^{[14]}$ predicted that in the certain kinematical
condition, inelastic neutrino-nucleus interaction should demonstrate
shadowing effect. Later, WA59 collaboration$^{[15]}$ compared the
kinematical distributions of neutrino and antineutrino events in the
neon and deuterium target experiments under the similar experimental
conditions. Their results showed that the neutrino and antineutrino
charged cross sections per nucleon in neon are relatively smaller
than those in deuterium at low $Q^2$. This is the first experimental
evidence of the shadowing effect in neutrino interactions, and is
consistent with the PCAC prediction.

\hspace{0.5cm}

The structure functions in the cross section formulas of DIS are
merely related to the quark densities. An essential point of the
quark-parton model is the universality of quark and gluon densities,
no matter they are measured in the electromagnetic current
interaction or the neutrino charge current or the neutrino neutral
current interactions. Therefore, the only discrepancy in the
neutrino and the charged lepton DIS is the shadowing effect, other
nuclear effects in the two cases should be consistent.

\hspace{0.5cm}

Although there is no individual neutrino experiment on EMC effect,
the differential cross sections and structure functions have been
measured in neutrino-nucleus experiments in CCFR$^{[16,17,18]}$ and
NuTeV $^{[19]}$ at Fermilab, and in CDHSW$^{[20]}$ and
CHORUS$^{[21]}$ at CERN. These experimental data would help us to
understand the nuclear effects in the neutrino-nucleus interaction
further.

\hspace{0.5cm}

The global analysis of nuclear parton distribution functions were
carried out by Eskola et al.$^{[22]}$, Hirai et al.$^{[23,24]}$ and
de Florian and Sassot$^{[25]}$, respectively. In those analysis, the
leading-order(LO)Dokshitzer-Gribov-Lipatov-Altarelli-Parisi(DGLAP)
evolution was done by the first two groups, while the
next-to-leading-order (NLO) evolution was performed by the third
group. In 1999, Eskola, Kolhinen, Ruuskanen and Salgado(EKRS)
suggested a set of nuclear parton distributions by using the
$F^A_2/F^D_2$ data in deep inelastic $lA$ collisions and the nuclear
Drell-Yan dilepton cross sections measured in $pA$ collisions. The
covered kinematical ranges were $10^{-6}\leq x\leq 1$ and
$2.25GeV^2\leq Q^2\leq10^{4}GeV^{2}$ for the nuclear targets from
deuteron to heavy ones. Their results agree very well with the
relevant EMC data and the E772 data at Fermilab$^{[26]}$. In 2001,
Hirai, Komano and Miyama(HKM)$^{[23]}$ proposed quadratic and cubic
types of nuclear parton distributions whose parameters were
determined by a $\chi^2 $ global fit to the available experimental
data, except those from the proton-nucleus Drell-Yan process. The
covered kinematical ranges were $10^{-9}\leq x\leq 1$ and
$1GeV^2\leq Q^2\leq10^{5}GeV^{2}$ for deuteron and heavy nuclear
targets. Their results reasonably explained the measured data of
$F_2$. In 2004, Hirai, Komano and Nagai(HKN)$^{[24]}$ re-analyzed
the measured ratios of nuclear structure functions
$F^{A}_{2}/F^{A'}_{2}$ and the ratios of Drell-Yan cross sections
between different nuclei for obtaining another parton distribution
function in nuclei. By employing the Drell-Yan data $^{[26,27]}$, as
well as the HERMES data $^{[28]}$, HKN determined the sea quark
modification in the range of $0.02<x_2<0.2$. It should be mentioned
that up to now no neutrino-nucleus DIS data have been included in
the analysis.

\hspace{0.5cm}

In this work, by means of the global LO DGLAP analyses of nuclear
parton distribution functions, the differential cross-sections and
the structure functions $F_2(x,Q^2)$ and $xF_3(x,Q^2)$ in the
neutrino-nucleus and anti-neutrino-nucleus charged-current DIS are
calculated and compared with the relevant data from Fermilab. It is
found that in the high and medium x regions, the anti-shadowing and
the EMC effects in the structure functions are the same, but the
nuclear corrections in the differential cross sections of
anti-neutrino charged current DIS are distinguishable. In sect.II, a
brief formulism for the differential cross section and the structure
function in the charged-current neutrino DIS is presented. The
result and discussion are given in sect.III, and the summary is
given in sect.IV.

\hspace{0.5cm}

{\bf II Brief formulism for differential cross section and structure
functions in charged-current neutrino DIS}

\hspace{0.5cm}

In the lab frame, the inclusive neutrino (anti-neutrino)-nucleon DIS
$^{[29-32]}$ can be described by three kinematic variables: the
squared momentum transfer $Q^2$, the incoming neutrino
(anti-neutrino) energy $E$, and the inelasticity variable $y$
representing the fractional energy transferred to the final hadronic
system.  $Q^2$ can be expressed in terms of the fraction $x$ of the
bound nucleon momentum,
\begin{equation}
  Q^2=2xyM_NE,
\end{equation}
where $M_N$ is the nucleon mass. If the parton mass is neglected,
both $x$ and $y$ are ranged from 0 to 1.

\hspace{0.5cm}In the single-W exchange approximation, the
differential cross sections for the charged-current
neutrino(anti-neutrino)-nucleus process in the very small final
lepton mass limit can be written as
\begin{equation}
 \frac{d\sigma^{\nu,\bar{\nu}}}{dxdy}=\frac{G_F^2EM_N}{\pi(1+Q^2/M^2_W)}
 [\frac{y^2}{2}2xF_1(x,Q^2)+(1-y-\frac{M_Nxy}{2E})F_2(x,Q^2)
 \pm y(1-\frac{y}{2})xF_3(x,Q^2)],
\end{equation}
where $G_F$ is the weak Fermi coupling constant, $M_W$ denotes the
mass of the W boson, and the + and - signs correspond to the $\nu$
and $\bar{\nu}$ scattering, respectively. In this equation, there
are three structure functions: $2xF_1(x,Q^2)$, $F_2(x,Q^2)$ and
$xF_3(x,Q^2)$. The first two structure functions are analogue to
those for charged-lepton DIS. The third structure function
$xF_3(x,Q^2)$ appears only in the weak interaction due to the
parity-violation term in the product of the leptonic and hadronic
tensors.

\hspace{0.5cm}

In order to account for the threshold correction of the heavy quark
production, a so-called slow re-scaling method is employed
$^{[33]}$. Then the structure function should be scaled by $\xi_S$,
rather than $x$,
\begin{equation}
\xi_S=x(1+\frac{m^2_k}{Q^2}),
\end{equation}
where $m_k$ is the heavy quark  mass with flavor $k$. The target
mass effect is further taken into account by evaluating quark
distributions at the Nachtmann variable $\xi_N$ $^{[34]}$ ,rather
than the Bjorken variable $x$:
\begin{equation}
\xi_N=\frac{2x}{1+\sqrt{1+4M^2_Nx^2/Q^2}}.
\end{equation}
At high $Q^2(Q^2\gg M^2_N)$, $\xi_N$ is equivalent to $x$. When the
target mass and heavy quark mass effects are simultaneously taken
into account, the Bjorken scaling variable $x$ should be replaced by
\begin{equation}
\xi_k=2x\frac{1+\frac{m^2_k}{Q^2}}{1+\sqrt{1+\frac{4M^2_Nx^2}{Q^2}(1+\frac{m^2_k}{Q^2})}}.
\end{equation}

\hspace{0.5cm}

In the quark-parton model, the structure functions are determined in
terms of the quark distribution functions $u(x,Q^2), d(x,Q^2),
s(x,Q^2), c(x,Q^2)$ and the gluon distribution function $g(x,Q^2)$,
which satisfy QCD $Q^2$-evolution equations. Using above mentioned
ingredients, the structure function $F_1(x,Q^2)$ in the neutrino
charged-current reaction can be written as
\begin{eqnarray}
F_1^{W^+p}(x,Q^2)&=&d(\xi_N,Q^2)|V_{ud}|^2
                   +d(\xi_c,Q^2)|V_{cd}|^2\theta(\xi_{Nc}-\xi_N)\nonumber\\
            & &    +\bar{u}(\xi_N,Q^2)(|V_{ud}|^2+|V_{us}|^2)
                   +\bar{u}(\xi_b,Q^2)|V_{ub}|^2\theta(\xi_{Nb}-\xi_N)\nonumber\\
            & &    +s(\xi_N,Q^2)|V_{us}|^2
                   +s(\xi_c,Q^2)|V_{cs}|^2\theta(\xi_{Nc}-\xi_N) \nonumber\\
            & &    +\bar{c}(\xi_N,Q^2)(|V_{cd}|^2+|V_{cs}|^2)
                   +\bar{c}(\xi_b,Q^2)|V_{cb}|^2\theta(\xi_{Nb}-\xi_N),
\end{eqnarray}
because the virtual $W^+$ coupled to the quarks with negative
charge. Similarly, the structure function $F_1(x,Q^2)$ in the
antineutrino charge-changing reaction can be expressed as
\begin{eqnarray}
F_1^{W^-p}(x,Q^2)&=&u(\xi_N,Q^2)(|V_{ud}|^2+|V_{us}|^2)
                   +u(\xi_b,Q^2)|V_{ub}|^2\theta(\xi_{Nb}-\xi_N)\nonumber\\
            & &    +\bar{d}(\xi_N,Q^2)|V_{ud}|^2
                   +\bar{d}(\xi_c,Q^2)|V_{cd}|^2\theta(\xi_{Nc}-\xi_N)\nonumber\\
            & &    +\bar{s}(\xi_N,Q^2)|V_{us}|^2
                   +\bar{s}(\xi_c,Q^2)|V_{cs}|^2\theta(\xi_{Nc}-\xi_N) \nonumber\\
            & &    +c(\xi_N,Q^2)(|V_{cd}|^2+|V_{cs}|^2)
                   +c(\xi_b,Q^2)|V_{cb}|^2\theta(\xi_{Nb}-\xi_N),
\end{eqnarray}
because the virtual $W^-$ coupled to the quarks with positive
charge. In these two equations, the quantities $V_{ij}$ are the
Cabibbo-Kobayashi-Maskawa(CKM) quark mixing matrix elements
$^{[35]}$, $\theta(\xi_{Nc}-\xi_N)$ and $\theta(\xi_{Nb}-\xi_N)$ are
step functions. The quantity $\xi_{Nk}$ can be defined as

\begin{equation}
\xi_{Nk}=\frac{Q^2}{Q^2+(M^{min}_X)^2-M^2_N},
\end{equation}
where $M^{min}_X$ is the minimum mass of the final hadron system for
the light quark transition to the heavy quark $k$.

\hspace{0.5cm}

The structure functions $F_2^{W^{\pm}p}(x,Q^2)$ and
$F_3^{W^{\pm}p}(x,Q^2)$ can be obtained from (6) and (7) by making
the replacements indicated in the curly brackets:
\begin{equation}
F_2^{W^{\pm}p}(x,Q^2)=F_1^{W^{\pm}p}\{q(\xi_N,Q^2) \rightarrow
             2xq(\xi_N,Q^2),q(\xi_k,Q^2) \rightarrow
             2\xi_kq(\xi_k,Q^2)\},
\end{equation}
\begin{equation}
F_3^{W^{\pm}p}(x,Q^2)=2F_1^{W^{\pm}p}\{\bar{q}(\xi_N,Q^2)
\rightarrow
             -\bar{q}(\xi_N,Q^2)\}.
\end{equation}
Assuming the isospin symmetry, corresponding neutron structure
functions can be obtained from the proton's by making replacements
$u(x,Q^2) \rightarrow d(x,Q^2)$ and $\bar{u}(x,Q^2)\rightarrow
\bar{d}(x,Q^2)$.

\hspace{0.5cm}

In the charged-lepton DIS,  the structure function $F_2(x,Q^2)$ is
related to the structure function $2xF_1(x,Q^2)$ by well-known
Callan-Gross relation$^{[36]}$. But, this relation is only valid if
the virtual photon is completely transverse. In fact, it has been
observed that Callan-Gross relation is not accurately held, and the
violation can usually be written as
\begin{equation}
2xF_1(x,Q^2)=\frac{1+4M^2x^2/Q^2}{1+R(x,Q^2)}F_2(x,Q^2),
\end{equation}
where $R(x,Q^2)$ is the ratio of the cross sections for the
longitudinally polarized photon to the transversely polarized
photon. An analogous relation should be held in the neutrino DIS.
The CHORUS $^{[21]}$ results on
 $R(x,Q^2)$ are in agreement with the more precisely measured values
in the charged-lepton scattering. By fitting the experimental data
available, Whitlow et al$^{[37]}$ gave an imperial expression
\begin{equation}
R(x,Q^2)=\frac{0.0635}{\log
(Q^2/0.04)}\theta(x,Q^2)+\frac{0.5747}{Q^2}-\frac{0.3534}{Q^4+0.09},
\end{equation}
where $\theta(x,Q^2)=1.0+\frac{12Q^2}{Q^2+1.0}\times
\frac{0.125^2}{0.125^2+x^2}$.

\hspace{0.5cm}

{\bf III Results and discussion }

\hspace{0.5cm}

Structure functions $F_2(x,Q^2)$ and $xF_3(x,Q^2)$ obtained from
neutrino scattering experiments are usually extracted from the sum
and the difference of the neutrino and the anti-neutrino y-dependent
differential cross sections, respectively. The structure
function$F_2(x,Q^2)$ can be expressed by the average of $F^{\nu
A}_2(x,Q^2)$ and $F^{\bar{\nu}A}_2(x,Q^2)$, while the structure
function $xF_3(x,Q^2)$ can be determined by $\frac{1}{2}(xF^{\nu
A}_3(x,Q^2)+xF^{\bar{\nu}A}_3(x,Q^2))$. In order to compare with the
experimental data, the expressions of $F_2(x,Q^2)$ and $xF_3(x,Q^2)$
are written as
\begin{equation}
F_2(x,Q^2)=\frac{1}{4}(F^{\nu p}_2(x,Q^2)+F^{\nu n}_2(x,Q^2)
           +F^{\bar{\nu}p}_2(x,Q^2)+F^{\bar{\nu}n}_2(x,Q^2)),
\end{equation}
\begin{equation}
xF_3(x,Q^2)=\frac{1}{4}(xF^{\nu p}_3(x,Q^2)+xF^{\nu n}_3(x,Q^2)
           +xF^{\bar{\nu}p}_3(x,Q^2)+xF^{\bar{\nu}n}_3(x,Q^2)).
\end{equation}

\hspace{0.5cm}

In our calculation,  the values of the CKM matrix elements are taken
from the global fit$^{[38]}$. They are $V_{ud}=0.9739$,
$V_{us}=0.221$, $V_{cd}=0.221$ and $V_{cs}=0.9730$. In the case of
heavy quark production, only those structure functions related to
the charm quark production are considered. The value of charm quark
mass is taken to be 1.31 GeV, which corresponds to the value
obtained in the LO QCD analysis of dimuon events$^{[39]}$. In terms
of the CTEQ (Coordinated Theoretical Experimental Project on QCD)
$^{[40]}$ parton distribution functions in proton and nuclear parton
distribution functions from EKRS$^{[22]}$ and HKN$^{[24]}$ (called
EKRS fit and HKN fit, respectively, in the rest of the paper), the
differential cross sections and the structure functions $F_2(x,Q^2)$
and $xF_3(x,Q^2)$ for charged-current neutrino and anti-neutrino
scatterings from iron are calculated. The results are plotted in
Figs.1-5, where the solid and the dashed curves represent the
results by using EKRS and HKN nuclear parton distributions with
nuclear effects, respectively, and the dotted curves denote the
results by employing CTEQ parton distributions without nuclear
effects. The theoretical result of the structure function
$F_2(x,Q^2)$ is compared with the NuTeV and CCFR experimental data
in Fig.1. In this figure, experimental data are taken from Ref.17
(open circle), Ref.18 (solid circle) and Ref.19 (open square),
respectively. It seems that the CHORUS results $^{[21]}$ favor the
CCFR data and the expected fact that the difference between the
nuclear structure functions of lead and iron is small. Preliminary
NuTeV data of $F_2(x,Q^2)$ $^{[19]}$ are generally consistent with
the CCFR data in the low and medium $x$ regions, but become larger
than the CCFR values when $x\geq0.65$. This deviation should be
confirmed in the further experiment. It is shown that our results
with nuclear effects agree excellently with the CCFR data in the
region of $x \geq 0.45$, which clearly shows the EMC effect in the
neutrino DIS. In the region of $0.14\leq x\leq0.35$, EKRS fit is in
a good agreement with the experimental data, but HKN fit apparently
overestimates the values of structure functions $F_2(x,Q^2)$. In the
smaller $x$ region, say $x=0.11$ or $x=0.09$, EKRS and HKN fits
reasonably describe the experimental data. They also show the
existence of anti-shadowing effect in the region of $0.09\leq
x\leq0.275$. In the very small $x$ region, say $x<0.07$, the
theoretical results apparently deviate from the the experimental
data with the decreasing value of $x$, especially in the $x<0.0175$
region. It is well-known that in the fixed target experiment, the
lower $x$ value usually corresponds to a low $Q^2$ value. The low
$x$ and low $Q^2$ structure function $F_2(x,Q^2)$ from neutrino
scattering experiments should not be necessary to agree with those
from the charged-lepton scattering experiment, because of the
contributions from the PCAC of the weak interaction. The shadowing
effect should be process-dependent, and should be smaller in the
neutrino DIS than in the charged-lepton DIS. So, the difference of
the structure function $F_2(x,Q^2)$ between the neutrino and the
charged-lepton reactions should further be investigated in the
experimental and the theoretical studies.

\hspace{0.5cm}

The behavior of the structure function $xF_3(x,Q^2)$ is presented in
Fig.2. In general, CHORUS results $^{[21]}$ are in agreement with
the CCFR and CDHSW$^{[20]}$ data. Preliminary NuTeV data are
consistent with the CCFR data in the low and medium $x$ regions, but
show higher values at $x\geq0.75$. Considering the nuclear effects,
our calculated results reasonably agree with the experimental data
in the region of $x \geq0.45$, showing the same nuclear effect
presented in the charged-lepton DIS. In the region of $0.18\leq x
\leq 0.35$, EKRS results agree with the experimental data, but HKN
results overestimate the structure function $xF_3(x,Q^2)$. In the
smaller $x$ region, say $x=0.11$ or $x=0.14$, EKRS and HKN fits
cannot properly describe the experimental data. Moreover, in the $x
\leq 0.09$ region, HKN fit is consonant with the data, but EKRS and
CTEQ fits overestimate the values of the structure function
$xF_3(x,Q^2)$. In  charged-lepton DIS, it is not obvious whether the
valence quark distribution indicate shadowing and anti-shadowing.
Nevertheless, the neutrino DIS experimental data expose the
shadowing and anti-shadowing effects in nuclear valence quark
distributions. In terms of the HKN nuclear parton distribution, one
can well describe the shadowing effect, but still overestimates the
anti-shadowing effect in the valence quark distribution.

\hspace{0.5cm}

The recombined experimental data and the structure function
$xF_3(x,Q^2)$ with various $Q^2$ are shown in the Fig.3. From this
figure, one sees that the HKN results consist excellently with the
experimental data at the low and high $x$ regions, but overestimate
the anti-shadowing effect of the valence quark in the bound nucleus
in the medium $x$ region with $Q^2\leq 50.1GeV^2$. Although EKRS fit
overestimate the values of structure functions at smaller $x$ in the
$Q^2\leq 12.6GeV^2$ region, it is in very good agreement with the
experimental data in the anti-shadowing and EMC effect regions with
$Q^2\geq20.0GeV^2$.

\hspace{0.5cm}

The differential cross sections for charged-current neutrino
(anti-neutrino) scattering on the iron nucleus are calculated. The
results are plotted and compared with experimental data in Fig.4. It
should be remarked that the NuTeV data $^{[19]}$ are reasonably in
agreement with the CCFR and CDHSW data with the exception of the
Bjorken variable  $x\geq0.40$ region, where CCFR neutrino and
anti-neutrino differential cross section data are unanimously below
the NuTeV results. The comparison with experimental data reveals
that the nuclear corrections are negligible in the differential
cross sections of charged-current anti-neutrino DIS, which is the
same as that in Ref.[41], but the neutrino-iron differential cross
sections provide more information. It is seen that there are not
nuclear effects in the region $x\leq0.08$. The results from EKRS and
CTEQ are in agreement with the experimental data, the HKN results
(greatly) overestimate the neutrino-nucleus differential cross
sections in the region of $0.125\leq x \leq 0.35$. This is because
that the sign of the $xF_(x,Q^2)$ term in the differential cross
section of neutrino DIS is positive, the contributions from valence
quarks are dominant. For $x \geq 0.45$, it seems that these three
fits overestimate the data a little bit.


Recently, S.A.Kulagin and H.Petti $^{[42]}$ (KP) and J.W.Qiu and
I.Vitev $^{[43]}$ (QV) respectively predicted the nuclear
corrections in the low $x$ region. Because HKN nuclear distribution
can provide a good description for structure function $xF_3(x,Q^2)$
at small $x$, we show the calculated ratios of $xF_3(Fe)$ to
$xF_3(D)$ with different $x$ and $Q^2$ values in Fig.5 and tabulate
them, as well as the KP's and QV's, in Table 1 for comparison. It is
shown that HKN fit present suppressions of $7\%$ and about $5\%$ at
$x=0.0001$ and $x=0.01$, respectively, while QV gave $15\%$ at both
$x=0.0001$ and $x=0.01$ with $Q^2=1.0GeV^2$, and KP showed a larger
nuclear suppression in the shadowing region.

\tabcolsep0.5cm
\begin{table}
\caption{The results in detail from HKN, QV and KP}
\begin{center}
\begin{tabular}{|c|c|c|c|c|c|c|}\hline
        & \multicolumn{3}{|c|}{HKN} & QV & \multicolumn{2}{|c|}{KP} \\
\hline
 $Q^2(GeV^2)$ & 1.0 & 5.0 & 20.0 & 1.0 & 5.0 & 20.0
\\ \hline
$x=10^{-4}$ & 0.929 & 0.931 & 0.933 & 0.85-0.88& 0.76 & 0.88
\\ \hline
$x=10^{-2}$ & 0.945 & 0.958 & 0.966 & 0.85-0.88 & 0.82 & 0.92
\\ \hline
\end{tabular}
\end{center}
\end{table}

\hspace{0.5cm}

In the global DGLAP analysis of nuclear effects, the abundant data
are used. They are the ratios of structure functions in the electron
and moun DIS' and the ratios of differential cross sections of the
lepton pairs production in the nuclear Drell-Yan process for
different nuclei, although there might exists the energy loss effect
$^{[44]}$.

The data of nuclear structure functions in neutrino-nucleus DIS are
so scarce that they have not been included in the current global fit
for nuclear parton distributions. Consequently, it is difficult to
determine nuclear valence quark distributions in the small x region
and nuclear anti-quark distributions in the $x>0.2$ region. In
contrast, the nuclear valence quark distribution in the medium and
large x regions can relatively be well determined. It is
conventionally considered that the EMC effect and Fermi motion
primarily occur in the scattering on valence quarks. The
anti-shadowing occurred in the medium $x$ region could be affected
by either the sea or the valence quark contributions. The shadowing
effect mainly comes from the scattering off the sea quark. The
nuclear modifications for neutrino and charged-lepton scatterings
should be expected to be identical without the lower $x$ and lower
$Q^2$ region due to the PCAC of the weak interaction. The
experimental data of $xF_3(x,Q^2)$ present obvious anti-shadowing
effect of valence quarks, which is absent in the charged-lepton DIS.

Therefore, it would be plausible if the neutrino DIS experimental
data can be included into the study of nuclear parton distributions.
In fact, by means of the structure function $xF_3(x,Q^2)$ in
neutrino DIS only, the nuclear modifications to the valence quark
distribution can very precisely be determined in the medium and
large x regions. With the structure functions $F_2(x,Q^2)$ from the
neutrino and charged-leptons scatterings, the nuclear modifications
to the sea quark distribution in the medium and large $x$ regions
would be pinned down. In addition, a detailed investigation of the
nuclear correction in the lower $x$ region is needed, because such a
correction depends on whether the process is the neutrino or the
charged-lepton DIS.

\hspace{0.5cm}

{\bf IV Concluding remarks }

\hspace{0.5cm}

As a summary, a LO analysis of neutrino-nucleus DIS is performed.
The structure functions $F_2(x,Q^2)$ and $xF_3(x,Q^2)$ and the
differential cross sections are calculated and compared with the
experimental data from CCFR and NuTeV by employing more appropriate
EKRS and HKN nuclear parton distributions and CTEQ parton
distributions without nuclear corrections. It is found that the
nuclear corrections are negligible in the differential cross
sections of the anti-neutrino charged-current DIS. The EMC effect
does exist in the neutrino structure function $F_2(x,Q^2)$. Such an
effect is as strong as that showed in the lepton structure function.
Shadowing and anti-shadowing effect occurs in $xF_3(x,Q^2)$ in the
small and  medium $x$ region,respectively. It clearly demonstrates
the shadowing and  anti-shadowing effect of the valence quark
distribution. Shadowing effects at small $x$ in the neutrino and the
lepton DIS are not exactly the same. This is due to the conservation
of the vector current in the lepton DIS and the PCAC of the weak
interaction. Shadowing effect in neutrino DIS should be weaker than
that in lepton DIS. Because of the process dependence of the
shadowing effect, further investigations are requested.  It is
necessary to measure the ratios of structure functions $F_2(x,Q^2)$
( and $xF_3(x,Q^2)$) for various nuclei in neutrino DIS. The
structure function $F_2(x,Q^2)$ is also very important to
investigate the nuclear shadowing effect in small x region by meams
of the charged-lepton DIS because the structure function ratios of
heavy nucleus to light nucleus are currently taking in lepton DIS
experiments. The MINERv-A(Fermilab E938)$^{[45]}$ and
neutrino-factory$^{[46]}$ projects will start in the near future.
The study of structure functions would deepen our knowledge of the
shadowing effect on valence quark and anti-quark distributions in
the neutrino and the lepton DIS processes. Clarifying the shadowing
effect will allow us to determine the nuclear modifications of
parton distributions, to study the new state of matter in the heavy
ion collision accurately, as well as to investigate the basic QCD
and electro-weak parameters.

{\bf Acknowledgement:}The authors thank D.Naples for the new NuTeV
experimental data by e-mail. This work is partially supported by
Natural Science Foundation of China(10475089,10435080,10575028) ,
CAS Knowledge Innovation Project (KJCX2-SW-N02),Major State Basic
Research Development Program (G20000774),  Natural Science
Foundation of Hebei Province(103143) and IHEP grant No.U529.

\newpage
\noindent\textbf{Figure caption}
\hspace{0.5cm}

Fig.1 The structure functions $F_2(x,Q^2)$ as a function of $Q^2$ at
various Bjorken variable $x$ from neutrino-iron deep inelastic
scattering. The experimental data are taken from Ref.17(open
circle), Ref.18(solid circle) and  Ref.19(open square),
respectively. The solid and dashed lines are the results from EKRS
and HKN nuclear parton distributions with nuclear effects,
respectively. The dotted lines are the results from CTEQ parton
distributions with no nuclear effects.

\hspace{0.5cm}

Fig.2 The structure functions $xF_3(x,Q^2)$ as a function of $Q^2$
at various Bjorken variable $x$ from neutrino-iron deep inelastic
scattering. The experimental data are taken from Ref.17(solid
circle)and  Ref.19(open square), respectively. The comments are the
same as Fig.1.

\hspace{0.5cm}

Fig.3 The structure functions $xF_3(x,Q^2)$ as a function of Bjorken
variable $x$ at various $Q^2$ from neutrino-iron deep inelastic
scattering. The experimental data are taken from Ref.17.The comments
are the same as Fig.1.

\hspace{0.5cm}

Fig.4 The differential cross sections in $x$ bins for neutrino(left)
and anti-neutrino (right) at $E=85GeV$. The experimental data are
taken from Ref.16. The comments are the same as Fig.1.

\hspace{0.5cm}

Fig.5 The ratios of  $xF_3(x,Q^2)$ for neutrino scattering off iron
and deuterium. The curves are drawn for $Q^2=1.0,5.0,20.0GeV^2$ from
HKN nuclear parton distributions.

\newpage
\begin{figure}
\epsfig{figure=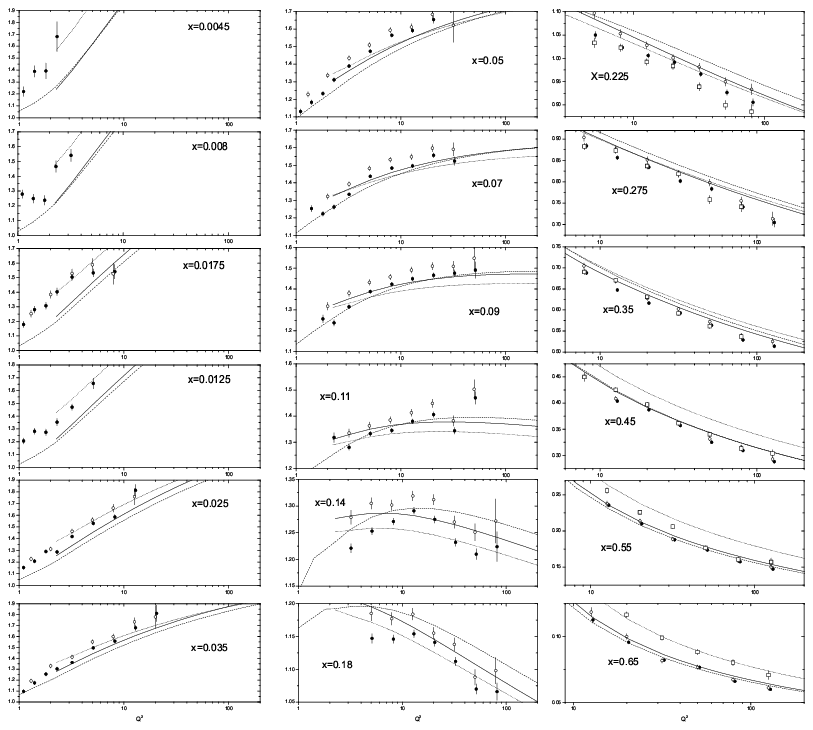,width=1.0\textwidth,height=20cm} \caption{}
\end{figure}

\newpage
\begin{figure}
\epsfig{figure=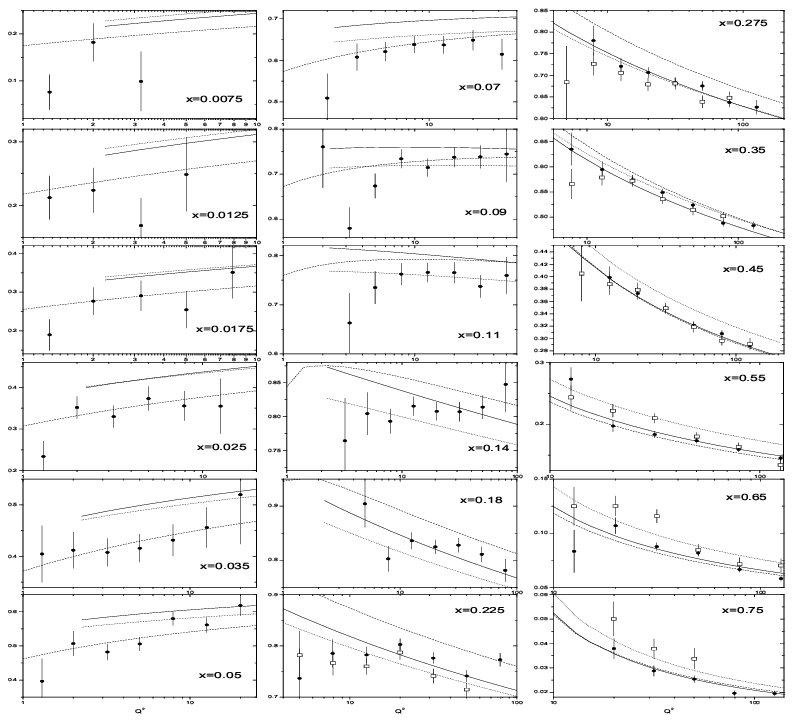,width=1.0\textwidth,height=23cm} \caption{}
\end{figure}

\newpage
\begin{figure}
\epsfig{figure=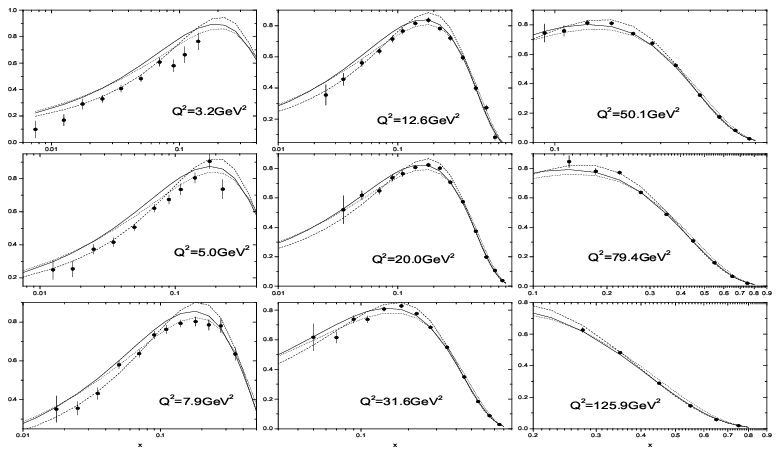,width=1.0\textwidth,height=23cm} \caption{}
\end{figure}

\newpage
\begin{figure}
\epsfig{figure=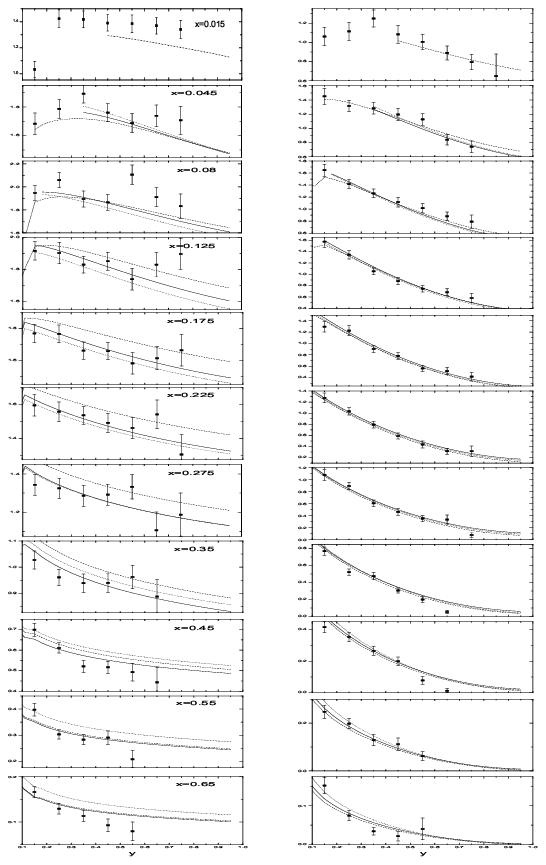,width=1.0\textwidth,height=23cm} \caption{}
\end{figure}

\newpage
\begin{figure}
\epsfig{figure=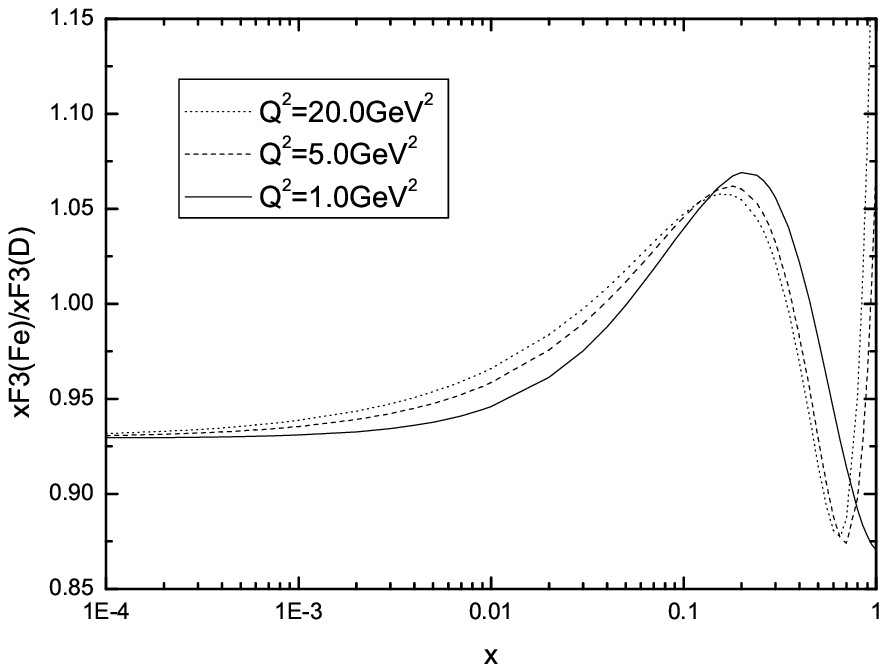,width=1.0\textwidth,height=17cm} \caption{}
\end{figure}


\vskip 1cm
\begin{thebibliography}{s2}
\bibitem{s1}  H.K.Ackermann et al.,Nucl.Instrum.Meth.A499(2003)624.
\bibitem{s2}  F.Carminati et al.,J.Phys.G30(2004)1517.
\bibitem{s3}  J.J.Aubert et al.,Phys.Lett.123B(1983)275.
\bibitem{s4} A.Bodek and J.L.Ritchie,Phys.Rev.D23(1981)1070;ibid,D24(1981)1400.
\bibitem{s5}  M.Arneodo,Phys.Rep.240(1994)301.
\bibitem{s6}  D.F.Geesaman,K.Saito,A.W.Thomas.Annu.Rev.Nucl.Part.Sci.45(1995)337.
\bibitem{s7}  G.Piller,W.Weise,Phy.Rep.330(2000)1.
\bibitem{s8}  P.R.Norton,Rep.Prog.Phys.66(2003)1253
\bibitem{s9}  A.M.Cooper et al.,Phys.Lett.B141(1984)133.
\bibitem{s10} H.Abramowicz et al(CDHS),Z.Phys.C25(1984)29.
\bibitem{s11} J.Hanlon.et al.,Phys.Rev.D32(1985)2441.
\bibitem{s12} J.Guy et al.,Z.Phys. C36(1987)337.
\bibitem{s13} S.L.Adler,Phys.Rev.B135(1964)963.
\bibitem{s14} J.S.Bell,Phys.Rev.Lett.13(1964)57.
\bibitem{s15} P.P.Allport et al.,Phys.Lett.B232(1989)417.
\bibitem{s16} U.K.Yang(CCFR),Ph.D.Thesis,University of
               Rochester,(2001),UR-1583.
\bibitem{s17} W.G.Seligman et al(CCFR),Phys.Rev.Lett.,79(1997)1213.
\bibitem{s18} B.T.Fleming,et al(CCFR),Phys.Rev.Lett.,86(2001)5430.
\bibitem{s19} M.Tzanov, et al(NuTeV),arXiv:hep-ex/0509010
\bibitem{s20} GOnengut et al(CHORUS), Phys.lett.B623(2006)65.
\bibitem{s21} J.P.Berge et al(CDHSW),Z.Phys.C49(1991)187.

\bibitem{s22} K.J.Eskola,V.J.Kolinen and C.A.Salgado(EKS),Eur.Phys.J.C9(1999)61.\\
              K.J.Eskola,V.J.Kolinen and P.V.Ruuskanen,Nucl.Phys.B535(1998)351.
\bibitem{s23} M.Hirai,S.Kumano,M.Miyama(HKM) ,Phys.Rev.D64(2001)034003.
\bibitem{s24} M.Hirai,S.Kumano,T.H.Nagai(HKN),Phys.Rev. C70 (2004) 044905, arXiv:hep-ph/0404093.
\bibitem{s25} D.de Florian and R.Sassot,Phys.Rev.D69(2004)074028.
\bibitem{s26} D.M.Adle et al.(E772),Phys.Rev.Lett.,64(1990)2479
\bibitem{s27} M.A.Vasiliev,et.al.(E866),Phys.Rev.Lett.83(1999)2304.
\bibitem{s28} A.Airapetian et.al.(HERMES),Phys.Lett.,B567(2003)339.
\bibitem{s29} S.R.Mishra and
              F.Sciulli,Annu.Rev.Nucl.Part.Sci.39(1989)269.
\bibitem{s30} M.Diemoz,F.Ferroni and
              E.Longo,Phys.Rept.,130(1986)293.
\bibitem{s31} J.M.Conrad, M.H.Shaevitz and
              T.Bolton,Rev.Mod.Phys.70(1998)1341.
\bibitem{s32} J.T.Londergan and A.W.Thomas Prog.Part.Nucl.Phys.41(1998)49.
\bibitem{s33} H.Georgi and H.D.Politzer,Phys.Rev.D14(1976)1829;\\
              R.M.Barnett,Physica D14(1976)70.
\bibitem{s34} O.Nachtmann, Nucl.Phys.B78(1974)455.
\bibitem{s35} M.Kobayashi and T.Maskawa,
              Prog.Theor.Phys.49(1973)652;\\
              N.Cabibbo,Phys.Rev.Lett.10(1963)531.
\bibitem{s36} C.G.Callan and D.G.Gross, Phys.Rev.Lett.22(1969)156.
\bibitem{s37} L.W.Whitlow et al.,Phys.lett.B250(1990)193.
\bibitem{s38} Particle Data Group,Phys.Lett.B592(2004)1.
\bibitem{s39} S.A.Rabinowitz et al.,Phys.Rev.Lett. 70(1993)134.
\bibitem{s40} H.L.lai et al.(CTEQ), Eur.Phys.J.C5(1998)461.
\bibitem{s41} J.J.Yang et al.,Phys.lett.,B546(2002)68.
\bibitem{s42} S.A.Kulagin and H.Petti, arXiv:hep-ph/0412425
\bibitem{s43}J.W.Qiu and I.Vitev, Phys.Lett.B587,(2004)52, arXiv:hep-ph/0401062
\bibitem{s44} C.G.Duan et al.,Eur.Phys.J.C39(2005)179;
     ibid,Eur.Phys.J.C.29(2003)557;
     Duan ChunGui,Wang HongMin and Li GuangLie,
     Chin.Phys.Lett.19(2002)485.
\bibitem{s45} MINERvA(E938),arXiv:hep-ex/0405002.
\bibitem{s46} S.Kumano, arXiv:hep-ph/0310166.
\end{thebibliography}
\end{document}